\documentclass[
twocolumn,
preprintnumbers,
nofootinbib,
prd,aps
]{revtex4}

\usepackage{epsf}
\usepackage{latexsym}
\usepackage{amssymb}
\usepackage{epsfig}
\usepackage{amsmath}
\usepackage{slashed}
\usepackage{lscape}

\begin{document}


\newcommand{\rem}[1]{$\spadesuit${\bf #1}$\spadesuit$}
\newcommand{\GeV}{{\text{GeV}}}
\newcommand{\eff}{{\text{eff}}}
\newcommand{\SM}{{\text{SM}}}
\newcommand{\BSM}{{\text{BSM}}}

\renewcommand{\topfraction}{0.8}

\preprint{UT-HET 106}

\title{
Gravitational waves as a probe of extended scalar sectors \\ with 
the first order electroweak phase transition }

\author{
Mitsuru Kakizaki, Shinya Kanemura, Toshinori Matsui
}
\affiliation{
Department of Physics, University of Toyama,
Toyama 930-8555, Japan
}


\begin{abstract}

We discuss spectra of gravitational waves which are originated by the strongly first order phase transition at the electroweak symmetry breaking, which is required for a successful scenario of electroweak baryogenesis.
Such spectra are numerically evaluated without high temperature expansion in a set of extended scalar sectors with additional $N$ isospin-singlet fields as a concrete example of renormalizable theories.
We find that the produced gravitational waves can be significant, so that they are detectable at future gravitational wave interferometers such as DECIGO and BBO.
Furthermore, since the spectra strongly depend on $N$ and the mass of the singlet fields, our results indicate that future detailed observation of gravitational waves can be in general a useful probe of extended scalar sectors with the first order phase transition.

\end{abstract}

\maketitle

\renewcommand{\thefootnote}{\#\arabic{footnote}}


\section{Introduction}

After the discovery of the Higgs boson ($h$) at LHC Run-I~\cite{Aad:2012tfa}, 
the standard model (SM) of elementary particles turned out to be a good description as the effective theory at the 
electroweak scale.
We have entered into a new stage to explore physics behind the Higgs sector, which can be related to the origin of phenomena beyond the SM, such as baryon asymmetry of the Universe (BAU)~\cite{Agashe:2014kda}, dark matter, cosmic inflation and neutrino oscillation.

Among various scenarios of BAU, electroweak baryogenesis (EWBG)~\cite{Kuzmin:1985mm} is directly connected with physics of the Higgs sector, requiring a strongly first order phase transition (1stOPT) at the electroweak symmetry breaking (EWSB) and also additional CP violating phases. 
It is known that new physics beyond the SM is necessary for EWBG. 
Such a scenario can be tested by experimental determination of the property of the Higgs sector.
For instance, the condition of the strongly 1stOPT can predict a significant deviation (order of several tens percent) in the triple Higgs boson coupling (the $hhh$ coupling) from the SM prediction~\cite{Kanemura:2004ch}, and the required CP violating phases lead to appearance of electric dipole moments, etc.

At the LHC experiment and its high luminosity one, 
the measurement of the $hhh$ coupling seems to be challenging. There is still a hope that in future the $hhh$ coupling could be measured by $13 \%$ accuracy~\cite{Asner:2013psa} at the upgraded version of the International Linear Collider (ILC). 

As a possible alternative method to test the strongly 1stOPT, we may be able to utilize future observation of gravitational waves (GWs)~\cite{Kamionkowski:1993fg}. 
Currently, GWs remain unobserved directly, and a number of observatories such as 
KAGRA~\cite{Somiya:2011np}, 
Advanced LIGO~\cite{Harry:2010zz}, Advanced VIRGO~\cite{Accadia:2009zz} are trying to detect them at first.
The target frequencies of GWs correspond to those from astronomical phenomena such as the binary of neutron stars, black holes, etc.. 
Once the GWs will be detected in the near future, the era of GW astronomy will come true.
Spectroscopy of GWs will make it possible to explore phenomena at the very early stage of the Universe, such as a strongly 1stOPT, cosmic inflation, topological defects like cosmic strings, domain wall, etc.

GWs originated from the strongly 1stOPT have been discussed in a model independent way in 
Refs.~\cite {Grojean:2006bp,Kikuta:2014eja,Espinosa:2010hh,No:2011fi,Hindmarsh:2015qta}.  
In the effective theory approach with higher order operators the possibility of detecting such GWs
was studied by Delaunay et al.~\cite{Delaunay:2007wb}.   
Apreda et al. evaluated spectra of GWs from the strongly 1stOPT due to thermal loop effects 
in the minimal supersymmetric 
SM (MSSM)~\cite{Apreda:2001us}, although such a scenario was already excluded by the LHC data.  
Espinosa et al. studied spectra of GWs in extended scalar sectors with the $O(N)$ 
symmetry~\cite{Espinosa:2007qk,Espinosa:2008kw}.
GWs from the non-thermal 1stOPT were investigated in singlet extensions of the SM~\cite{Ashoorioon:2009nf} and the MSSM~\cite{Apreda:2001us} and in the left-right symmetric model~\cite{Sagunski:2012ufa}.

In this paper, we discuss the possibility that  future detailed observation of GWs is useful not only to test the electroweak 1stOPT but also as a probe of extended scalar sectors and further the physics behind. 
To this end, we evaluate spectra of GWs from the strongly 1stOPT at the EWSB in a set of extended scalar sectors with additional $N$ isospin-singlet fields as an example of renormalizable theories which can cause the 1stOPT thermally. 
We find that the relic density of the produced GWs can be so significant that they are detectable at future GW interferometers such as DECIGO~\cite{decigo} and BBO~\cite{Corbin:2005ny}.
The spectra depend on $N$ and the mass of the additional scalar fields.
We conclude that GWs can be a useful probe of physics behind the Higgs sector.

\section{O(N) scalar singlet model}

We consider a set of extensions of the SM with additional $N$ isospin-singlet scalars 
$\vec{S}=(S_1, S_2, \cdots, S_N)^T$ invariant under an $O(N)$ symmetry, 
\begin{align}
V_0(\Phi, \vec{S})= V_{\rm SM}(\Phi)
+\frac{\mu_S^2}{2}|\vec{S}|^2 
+\frac{\lambda_S}{4}|\vec{S}|^4
+\frac{\lambda_{\Phi S}}{2}|\Phi|^2|\vec{S}|^2,  \nonumber
\end{align}
where $V_{\rm SM}$ is the Higgs potential of the SM.
After the EWSB, the SM Higgs doublet is parametrized as
\begin{align}
\Phi=
\begin{pmatrix} \omega^+ \\ \frac{1}{\sqrt{2}}(v+h+iz) \end{pmatrix},\nonumber
\end{align}

\begin{widetext}

\begin{figure}[t]

\begin{minipage}{0.38\hsize}
  \centerline{\epsfxsize=1\textwidth\epsfbox{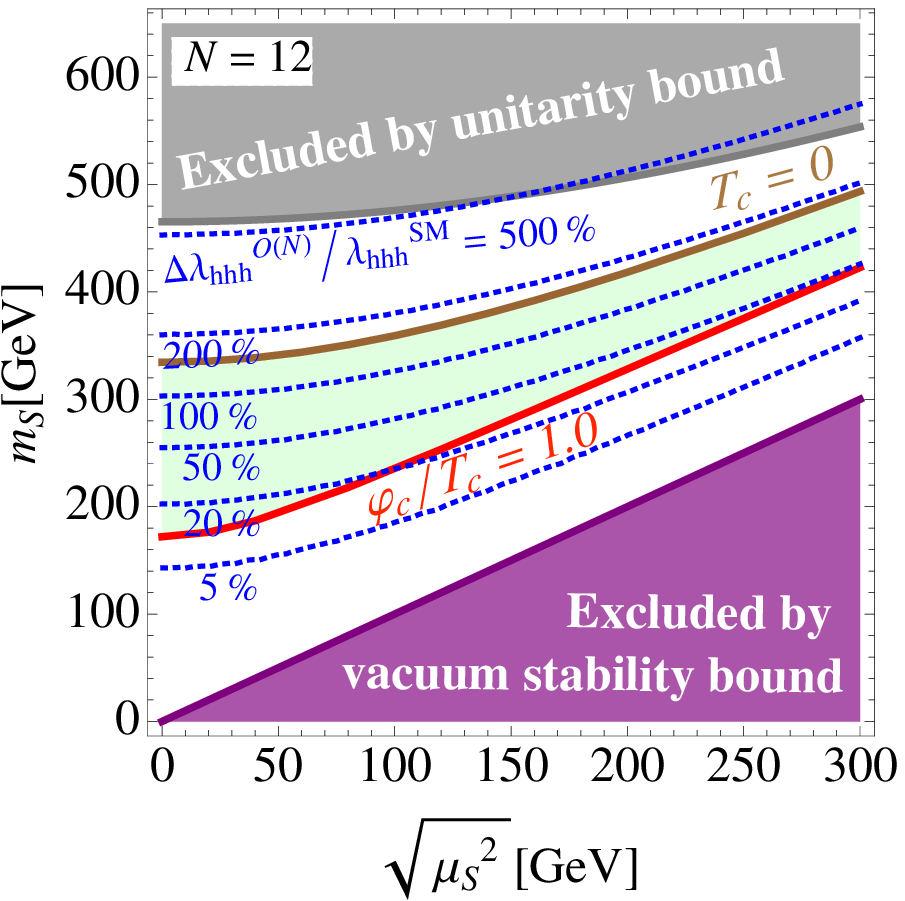}}
\end{minipage}
\hspace{2cm}
 \begin{minipage}{0.38\hsize}
  \centerline{\epsfxsize=1\textwidth\epsfbox{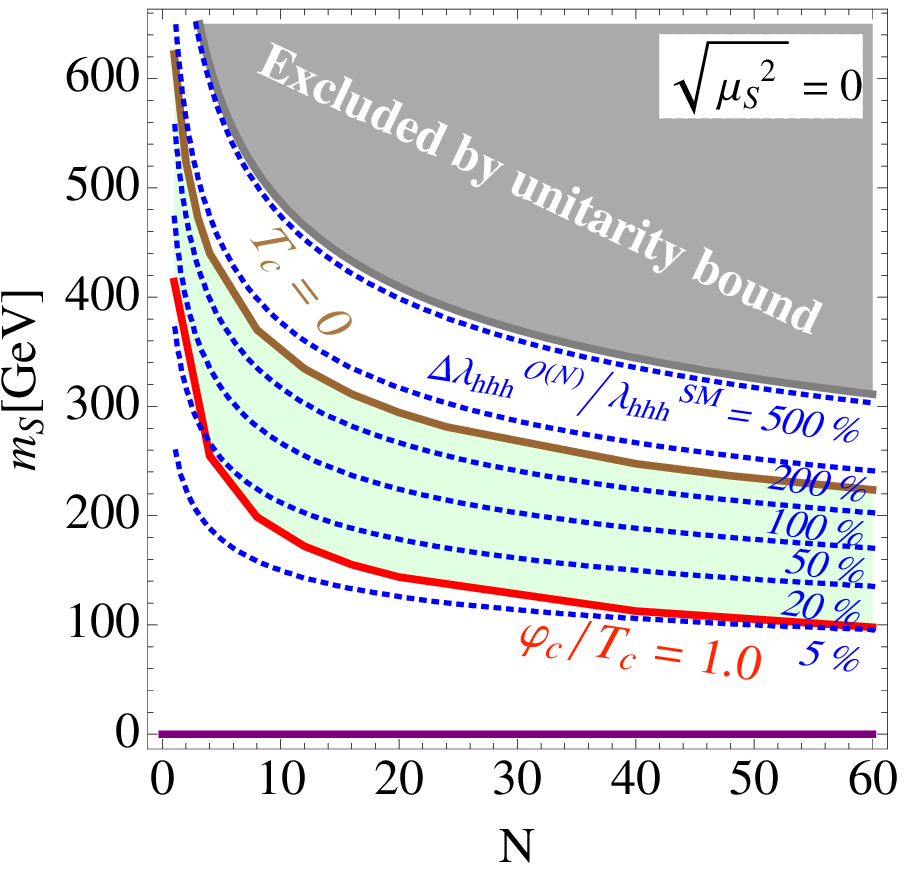}}
\end{minipage}
\caption{
The allowed region which satisfies both $\varphi_c/T_c> 1$ and $T_c>0$, where EWBG can be viable with the strongly 1stOPT on the plane of $\sqrt{\mu_S^2}$ and $m_S$ in the left figure and on the plane of $N$ and $m_S$ in the right figure.
We set $N=12$ for the left figure, and $\mu_S^2=0$ for the right figure.
Contours for the deviation in the $hhh$ coupling from the SM prediction are also shown in both figures.
Bounds from vacuum stability and perturbative unitarity are also shown for $\lambda_S^{} =0$~\cite{kkm_full}.
}
\label{fig1}
\end{figure}

\end{widetext}

\noindent
where $\omega^\pm$ and $z$ are Nambu-Goldstone bosons and $v$ ($\simeq 246$ GeV) 
is the vacuum expectation value (VEV). 
The $O(N)$ symmetry is assumed not to be spontaneously broken.
The mass of $h$ is set as  $m_h=125$ GeV, and the common 
mass of $S_i$ is given at the tree level by
\begin{align}
m_{S}^2=\mu_S^2+\frac{\lambda_{\Phi S}}{2}v^2. 
\label{masses}
\end{align}
We take $m_S^{}$, $\mu_S^2$ and $\lambda_S^{}$ as input free parameters 
in the scalar sector. 

The effective potential at finite temperatures is  given at the one-loop level by 
\begin{align}
V_{\eff}(\varphi,T)=V_{0}(\varphi) + \Delta V_1(\varphi)+ \Delta V_T(\varphi, T), 
\label{Eq.effpot}
\end{align}
where $\Delta V_1$ is the one-loop contributions with the counter term at zero temperature, 
in which field dependent masses of the gauge bosons in the loop are replaced by thermally corrected ones in Ref.~\cite{Carrington:1991hz}. That of the singlet scalars is replaced by
\begin{align}
\!\!\!\!M_S^2(\varphi) \to M_S^2(\varphi, T)=(m_S^2-\mu_S^2) \frac{\varphi^2}{v^2}+\mu_S^2+\Pi_S(T), \nonumber
\end{align}
where 
$\Pi_S(T) =\frac{T^2}{12v^2}[(N+2)\lambda_S v^2+4(m_S^2-\mu_S^2)]$.
The term $\Delta V_T$ is composed of thermal loop contributions~\cite{Dolan:1973qd}. 

We define the renormalized VEV $v$ and the renormalized mass $m_h$ 
of $h$ by the following two conditions,
\begin{align}
\frac{\partial V_{\eff}(\varphi, 0)}{\partial \varphi}\bigg|_{\varphi=v}
=0, \quad
\frac{\partial^2 V_{\eff}(\varphi, 0)}{\partial \varphi^2}\bigg|_{\varphi=v}
=m_h^2. \nonumber
\end{align}
The renormalized $hhh$ coupling at $T=0$ is defined by
\begin{align}
\lambda_{hhh} \equiv \left. \frac{\partial^3}{\partial \varphi^3}V_{\eff}(\varphi, 0) \right|_{\varphi=v}, \nonumber
\end{align}
and is calculated at the one loop level in our model as 
\begin{align}
\lambda_{hhh}^{O(N)}
=\frac{3m_h^2}{v}
\left\{1
-\frac{1}{\pi^2}\frac{m_t^4}{v^2m_h^2}
+\frac{N}{12\pi^2}\frac{m_S^4}{v^2m_h^2}\left(1-\frac{\mu_S^2}{m_{S}^2}\right)^3
\right\}.
\label{hhh_O(N)}
\end{align}
There are two sources for the physical common mass $m_S$ of the scalar fields $S_i$, as shown in Eq.~(\ref{masses}).
If $m_S$ is large because of a large value of $\mu_S$, the one loop correction in Eq.~(\ref{hhh_O(N)}) decouples in the large mass limit.
Instead, if $\mu_S$ is relatively small as $v$, the one loop contribution does not decouple and a quartic 
powerlike contribution for the mass remains in $\lambda_{hhh}^{O(N)}$~\cite{Kanemura:2002vm}.

One of the necessary conditions~\cite{Sakharov:1967dj} to generate BAU is the departure from thermal equilibrium. 
To satisfy this condition, the baryon number changing sphaleron interaction must quickly decouple 
in the broken phase, which is described by  
 $\Gamma(T)\lesssim H(T)$,
where $\Gamma(T)$ is the reaction rate of the sphaleron process and $H(T)$ is the Hubble parameter at $T$.
The above condition leads to a strongly 1stOPT, which is typically described by~\cite{Kuzmin:1985mm}
\begin{align}
\frac{\varphi_c}{T_c} \gtrsim 1,
\label{ft}
\end{align}
where $\varphi_c$ gives the broken phase minimum at the critical temperature $T_c$.
In this paper, we calculate $\varphi_c/T_c$ numerically without using high temperature expansion 
by using the ring-improved finite temperature effective potential in Eq.~(\ref{Eq.effpot}).

We show the region which satisfies both $\varphi_c/T_c> 1$ and $T_c>0$, where EWBG can be viable with the strongly 1stOPT on the plane of $\sqrt{\mu_S^2}$ and $m_S$ in Fig.~\ref{fig1} (left) and on the plane of $N$ and $m_S$ in Fig.~\ref{fig1} (right).
In Fig.~\ref{fig1} (left), we show the results for $N=12$.
In Fig.~\ref{fig1} (right), to obtain maximal non-decoupling effects, we set $\mu_S^2$ to be 0.
We also show contour plots for the deviation in the $hhh$ coupling 
from the SM prediction.

We find that, as indicated in Ref.~\cite{Kanemura:2004ch} in the case of the two Higgs doublet model (2HDM), 
significant deviations in the $hhh$ coupling appear in the allowed region of the strongly 1stOPT.
Notice that the scenario of the 2HDM in Ref.~\cite{Kanemura:2004ch} 
corresponds to $N=4$ in our model~\cite{Gil:2012ya}.
We emphasize that the correlation between the strongly 1stOPT and the large deviation in the $hhh$ coupling is a common feature of the models where the condition of quick sphaleron decoupling is satisfied by the thermal loop effects of additional scalar bosons.
This property can be utilized to test scenarios of EWBG by measuring the $hhh$ coupling at 
the ILC as we already pointed out. 

\section{Spectra of Gravitational Waves}

The relic abundance of GWs from the electroweak 1stOPT is composed of the contributions from 
bubble collisions and the turbulence in the plasma as~\cite{Grojean:2006bp}
\begin{align}
\Omega_{\rm GW} (f) h^2
= \Omega_{\rm coll} (f) h^2 +\Omega_{\rm turb} (f) h^2. \nonumber 
\end{align}
In our analysis, we employ the results of Ref.~\cite{Huber:2008hg} for the bubble collision contribution
\begin{align}\hspace*{-2mm}
\Omega_{\rm coll} (f) h^2 
&=\widetilde{\Omega}_{\rm coll} h^2 \times
  \begin{cases}
    \left(\frac{f}{\tilde{f}_{\rm coll}}\right)^{2.8} &\text{(for $f<\tilde{f}_{\rm coll}$)} \\
    \left(\frac{f}{\tilde{f}_{\rm coll}}\right)^{-1} &\text{(for $f>\tilde{f}_{\rm coll}$)}
  \end{cases},  \nonumber
  \end{align}
where the energy density is obtained as 
\begin{align}
\widetilde{\Omega}_{\rm coll} h^2
\simeq c \kappa^2 \left(\frac{H_t}{\beta}\right)^2 \left(\frac{\alpha}{1+\alpha}\right)^2 
\left(\frac{v_b^3}{0.24+v_b^3}\right)\left(\frac{100}{g^t_\ast}\right)^{1/3}, \nonumber
\end{align}
with  $c=1.1 \times 10^{-6}$ at the peak frequency given by 
\begin{align} \hspace*{-4mm}
\tilde{f}_{\rm coll} \simeq 5.2 \times 10^{-3} \mbox{mHz} \left(\frac{\beta}{H_t}\right)\left(\frac{T_t}{100 \mbox{GeV}}\right)\left(\frac{g^t_\ast}{100}\right)^{1/6}. \nonumber
\end{align}
For the plasma turbulence contribution, we use~\cite{Nicolis:2003tg} 
\begin{align}\hspace*{-2mm}
\Omega_{\rm turb} (f) h^2
&=\widetilde{\Omega}_{\rm turb} h^2 \times 
  \begin{cases}
    \left(\frac{f}{\tilde{f}_{\rm turb}}\right)^{2}     & \text{(for $f<\tilde{f}_{\rm turb}$)} \\
    \left(\frac{f}{\tilde{f}_{\rm turb}}\right)^{-3.5} &  \text{(for $f>\tilde{f}_{\rm turb}$)}
  \end{cases},  \nonumber
\end{align}
where the energy density is evaluated as 
\begin{align}
\widetilde{\Omega}_{\rm turb} h^2 \simeq 1.4 \times 10^{-4} u_s^5v_b^2 \left(\frac{H_t}{\beta}\right)^2 \left(\frac{100}{g^t_\ast}\right)^{1/3}, \nonumber
\end{align}
at the peak frequency given by
\begin{align}
\tilde{f}_{\rm turb} \simeq 3.4 \times 10^{-3} \mbox{mHz} \frac{u_s}{v_b}\left(\frac{\beta}{H_t}\right)\left(\frac{T_t}{100 \mbox{GeV}}\right)\left(\frac{g^t_\ast}{100}\right)^{1/6}. \nonumber
\end{align}
The bubble wall velocity $v_b(\alpha)$, the turbulent fluid velocity $u_s(\alpha)$ and the efficiency factor $\kappa(\alpha)$ are given in Ref.~\cite{Kamionkowski:1993fg}, and 
$g^t_\ast$ ($=g_\ast(T_t)$) is the total number of effective degree of freedom at 
the transition temperature $T_t$.  
$H_t$ is the Hubble parameter at $T_t$ in 
the radiation dominant Universe. 

The parameter $\alpha$ is the ratio of the false-vacuum energy density $\epsilon(T)$ and the thermal energy density $\rho_{\rm rad}(T)$ in the symmetric phase by 
\begin{align}
\alpha
\equiv \frac{\epsilon(T_t)}{\rho_{\rm rad}(T_t)} \nonumber
\end{align}
and
\begin{align}
\epsilon(T)
\equiv -\Delta V_{\eff}(\varphi_B(T),T)+T\frac{\partial \Delta V_{\eff}(\varphi_B(T), T)}{\partial T}, 
\nonumber
\end{align}
where $\Delta V_{\eff}(\varphi(T),T)$ is the free energy density with respect to that of the symmetric phase,
and $\varphi_B(T)$ is the broken phase minimum at $T$.
The radiation energy density is given by
$\rho_{\rm rad}(T)
=(\pi^2/30) g_*(T)T^4$. 

The parameter $\beta$ is defined  as
\begin{align}
\beta
\equiv
-\frac{d S_E}{d t}\bigg|_{t=t_t}
\simeq\frac{1}{\Gamma}\frac{d \Gamma}{d t}\Bigg|_{t=t_t}, \nonumber
\end{align}
where $t_t$ is the phase transition time, $S_E(T)\simeq S_3(T)/T$ with $S_3$ being the three dimensional Euclidean action, 
\begin{align}
 S_3 \equiv \int d^3r \left[\frac{1}{2}  (\vec{\nabla}\varphi)^2+ V_{\rm eff}(\varphi, T)\right], \nonumber
 \end{align}
and $\Gamma=\Gamma_0(T)\exp[-S_E(T)]$ is the rate of variation of the bubble nucleation rate 
with $\Gamma_0(T)\propto T^4$.
We then obtain the normalized dimensionless parameter as 
\begin{align}
\tilde{\beta} \equiv 
\frac{\beta}{H_t} = T_t\frac{d}{d T}\left(\frac{S_3(T)}{T}\right)\Bigg|_{T=T_t}. \nonumber
\end{align}
When the phase transition is complete; i.e.,
\begin{align}
\left.\frac{\Gamma}{H^4}\right|_{T=T_t} \simeq 1, 
\label{eq:G/H}
\end{align}
we obtain
$S_3(T_t)/T_t
=4\ln (T_t/H_t) 
\simeq 140-150$.

\begin{widetext}

\begin{figure}[t]
 \begin{minipage}{0.38\hsize}
  \centerline{\epsfxsize=1\textwidth\epsfbox{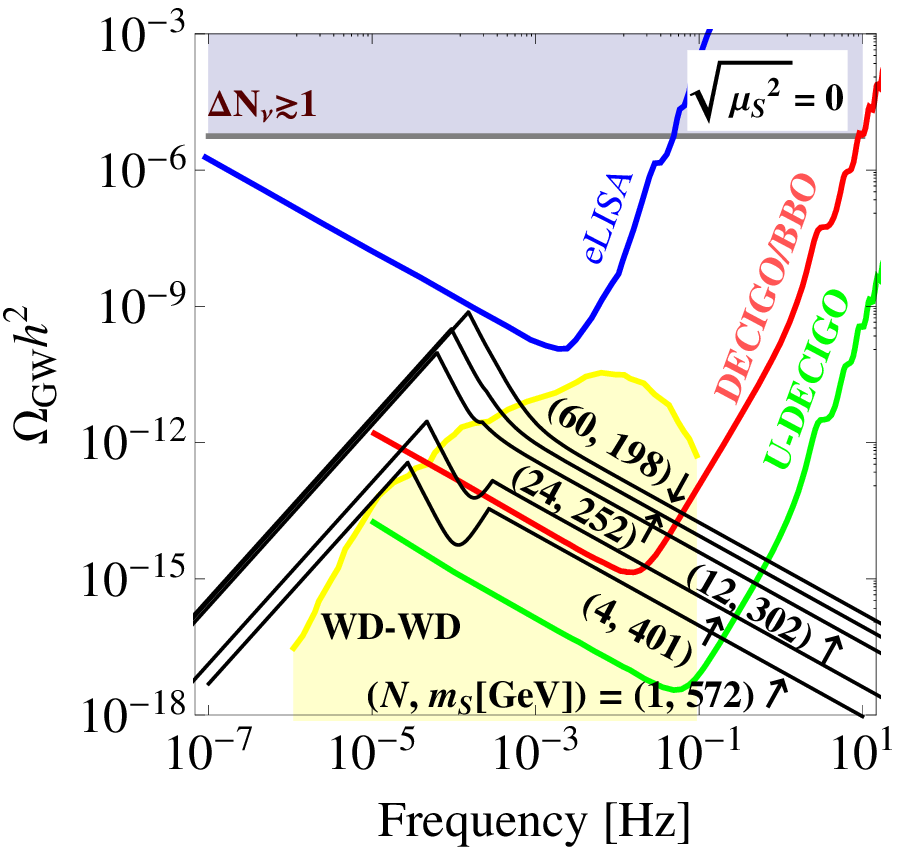}}
\end{minipage}
\hspace{2cm}
\begin{minipage}{0.38\hsize}
  \centerline{\epsfxsize=1\textwidth\epsfbox{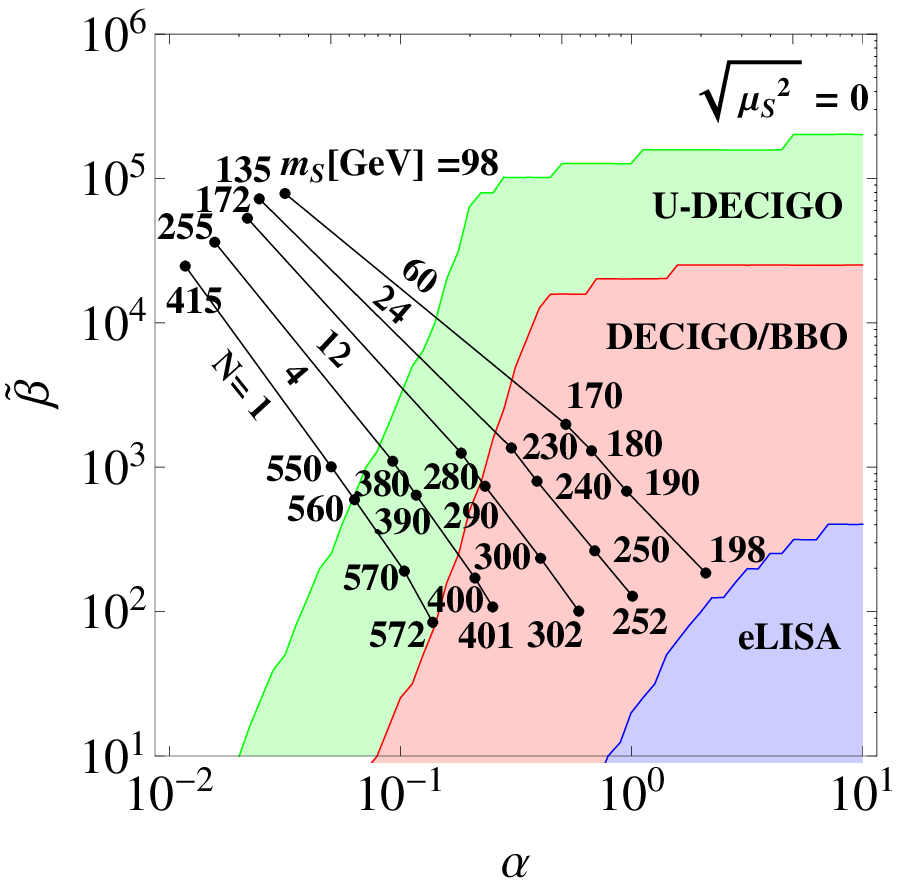}}
\end{minipage}
\caption{
(Left) Spectra of GWs in the $O(N)$ singlet model with expected experimental sensitivities 
at the future GW interferometers, eLISA, DECIGO/BBO and Ultimate-DECIGO (U-DECIGO) are shown for $\sqrt{\mu_S^2} =0$.  
The bound from non-observation of the energy density 
of extra radiation is indicated by $\Delta N_\nu \gtrsim 1$~\cite{planck, Agashe:2014kda}, and the estimated foreground noise from 
the white dwarf binaries is also shown. 
(Right) Predictions of the model on the $(\alpha, \tilde{\beta})$ plane with various $N$ and $m_S^{}$ 
assuming $\sqrt{\mu_S^2} =0$ and $T_t = 100$ GeV are shown with regions of expected experimental sensitivity at the future GW interferometers.  
}\label{fig2}
\end{figure}

\end{widetext}

In Fig.~\ref{fig2} (left), the predicted spectra of GWs are shown as a function of the frequency 
for $N=1$, $4$, $12$, $24$ and $60$ with $\sqrt{\mu_s^2} =0$ in the $O(N)$ singlet model. 
For each $N$, $m_S$ is taken its maximal value 
under the condition of the complete phase transition given in Eq.~(\ref{eq:G/H}).  
These sets of $(N, m_S)$ are all in the allowed region shown in Fig.~\ref{fig1}, 
where EWBG is possible.  
Curves of expected experimental sensitivities for GWs 
at eLISA, DECIGO/BBO and Ultimate-DECIGO are also shown~\cite{Seoane:2013qna,Kudoh:2005as}.  
Estimated foreground noise from white dwarf binaries in Ref.~\cite{Schneider:2010ks} are also shown. 
One can see that for larger $N$ the strength of GWs is more significant and 
the spectra are within the observable reach of DECIGO/BBO. 
Even for smaller values of $m_S$ or for the case of $N=1$, the  spectra  may be  
able to be observed at Ultimate-DECIGO. 

\noindent
There is a strong correlation between the strength of the GWs and the value of 
$\varphi_c/T_c$ (hence, $\Delta \lambda_{hhh}^{O(N)}/\lambda_{hhh}^{\SM}$).  

In Fig.~\ref{fig2} (right), we show the predictions of the model for $N=1$, $4$, $12$, $24$ and $60$ 
with various $m_S$ with $\sqrt{\mu_s^2} =0$ on the $(\alpha, \tilde{\beta})$ plane 
under the conditions of $\varphi_c/T_c >1$ and the complete phase transition.  
We set $T_t =100$ GeV, as the result is not very sensitive to $T_t$.
Regions of expected experimental sensitivity at eLISA, DECIGO/BBO and Ultimate-DECIGO 
are also shown.  
One can see that different sets of $(N, m_S)$ corresponds to different points on the $(\alpha, \tilde{\beta})$ plane. 
Therefore, future GW observation experiments can be a probe of 
distinguishing various models of the electroweak 1stOPT.   

\section{Conclusion}

We have investigated spectra of GWs which come from the strongly electroweak 1stOPT, which is required for a successful scenario of EWBG in a set of extended scalar sectors with additional $N$ isospin-singlet fields 
as a concrete example of renormalizable theories.  
The $hhh$ coupling also has been evaluated at the one loop level in these models.
The produced GWs can be significant, so that they are detectable at future GW 
interferometers such as DECIGO and BBO.
Furthermore, since the spectra strongly depend on $N$ and $m_S^{}$, 
we conclude that future detailed observation of GWs can be generally useful as a probe of extended scalar sectors with the 1stOPT. 
The detailed analyses are shown elsewhere~\cite{kkm_full}.

This work was supported by 
Grant-in-Aid for Scientific Research, No.\ 26104702 (MK) and No.\ 23104006 (SK), 
Grant H2020-MSCA-RISE-2014~no.~645722 (Non Minimal Higgs) (SK), 
and the Sasakawa Scientific Research Grant from The Japan Science Society (TM).

\end{document}